# Observation time dependent mean first passage time of diffusion and sub-diffusion processes

Ji-Hyun Kim[1], Hunki Lee[1,2], Sanggeun Song[1,2], Hye Ran Koh*[2], and Jaeyoung Sung*[1,2]

[1] *Creative Research Initiative Center for Chemical Dynamics in Living Cells, Chung-Ang University, Seoul 06974, Republic of Korea*

[2] *Department of Chemistry, Chung-Ang University, Seoul 06974, Republic of Korea*

**Author Information** The authors declare no competing financial interests. Correspondence and requests for materials should be addressed to J.S. (jaeyoung@cau.ac.kr) and H.R.K. (hrkoh@cau.ac.kr).

**Abstract**

The mean first passage time, one of the important characteristics for a stochastic process, is often calculated assuming the observation time is infinite. However, in practice, the observation time, $T$, is always finite and the mean first passage time (MFPT) is dependent on the length of the observation time. In this work, we investigate the observation time dependence of the MFPT of a particle freely moving in the interval $[-L,L]$ for a simple diffusion model and four different models of subdiffusion, the fractional diffusion equation (FDE), scaled Brown motion (SBM), fractional Brownian motion (FBM), and stationary Markovian approximation model of SBM and FBM. We find that the MFPT is linearly dependent on $T$ in the small $T$ limit for all the models investigated, while the large-$T$ behavior of the MFPT is sensitive to stochastic properties of the transport model in question. We also discuss the relationship between the observation time, $T$, dependence and the travel-length, $L$, dependence of the MFPT. Our results suggest the observation time dependency of the MFPT can serve as an experimental measure that is far more sensitive to stochastic properties of transport processes than the mean square displacement.

**Keywords:** First passage time, observation time dependence, fractional diffusion equation, scaled Brownian motion, fractional Brownian motion

## 1. Introduction

Diffusion is one of few universally observed phenomena in nature; thus, it has been thoroughly investigated in non-equilibrium statistical mechanics [1]. However, anomalous subdiffusion has also been observed across a variety of systems such as charge carrier transport in amorphous semiconductors [2-8], transport on fractal geometries [9, 10], the diffusion of a scalar tracer in an array of convection rolls [11, 12], the dynamics of a bead in a polymeric network [13, 14], or polymeric systems [15-18]. Subdiffusion is usually characterized by the following lag time-dependence of the mean square displacement:

$$\langle x^2(t) \rangle = \frac{2D_\alpha}{\Gamma(1+\alpha)} t^\alpha \quad (0 < \alpha < 1), \tag{1}$$

where $D_\alpha$ and $\Gamma(z)$ respectively denote a generalized diffusion constant and a gamma function defined by $\Gamma(z) = \int_0^\infty dt\, e^{-t} t^{z-1}$.

The fractional diffusion equation (FDE) is one of well-known subdiffusive transport equations [19, 20]

$$\frac{\partial}{\partial t} p(x,t) = {}_0 D_t^{1-\alpha} D_\alpha \frac{\partial^2}{\partial x^2} p(x,t), \tag{2}$$

where $p(x,t)$ denotes the probability distribution that a particle is located at position $x$ at time $t$. Equation (2) was derived by considering the continuous limit dynamics of the continuous-time random walk (CTRW) on a lattice characterized by a waiting time distribution, $\psi(t)$, of a random walker jumping to nearest neighboring sites with a heavy power-law tail, $\psi(t) \sim t^{-1-\alpha}$ $(0 < \alpha < 1)$ [2, 21-25]. In equation (2), $_0D_t^{1-\alpha}$ denotes the Riemann-Liouville fractional differential operator defined by

$$_0D_t^{1-\alpha} p(t) = \frac{1}{\Gamma(\alpha)} \frac{\partial}{\partial t} \int_0^t d\tau \frac{p(\tau)}{(t-\tau)^{1-\alpha}}. \tag{3}$$

For the FDE model, it is known that the mean first passage time (MFPT) does not exist, given that our observation time is infinite [26-28]. In experiments or simulations, however, observation time is always finite and we observe only those first passage events occurring in our observation time. When our observation time is finite, the average of such first passage times is also finite and dependent on the length, $T$, of our observation time. This observation time-dependent or $T$-dependent MFPT is defined as

$$\langle t \rangle_T = \int_0^T dt \, t f_T(t), \tag{4}$$

where $f_T(t)$ denotes the observation time-dependent first passage time distribution, defined by $f_T(t) \equiv f_\infty(t) \big/ \int_0^T dt f_\infty(t)$ $(0 < t < T)$ [29], where $f_\infty(t)$ denotes the first passage time distribution in the limit of infinite observation time, i.e., $f_\infty(t) = \lim_{T \to \infty} f_T(t)$. The $T$-dependent MFPT given in equation (4) can be represented in terms of the survival probability, defined by $S(t) = \int_t^\infty dt' f_\infty(t')$:

$$\langle t \rangle_T = \frac{\int_0^T dt \, S(t) - T S(T)}{1 - S(T)} \tag{5}$$

In this work, we investigate how the $T$-dependent MFPT for the FDE model compares to the $T$-dependent MFPTs for simple diffusion and various other subdiffusion processes.

## 2. Transport models

### 2.1. Diffusion equation

Let us consider a particle freely moving along the *x*-axis in the interval $(-L, L)$. The particle's initial position is chosen to be the coordinate origin for the sake of simplicity. Here, a first passage time is defined as a time taken for the particle to reach $-L$ or $L$ for the first time. When a particle's motion follows the classical diffusion equation, which is given by

$$\frac{\partial}{\partial t} p(x,t) = D_1 \frac{\partial^2}{\partial x^2} p(x,t) \tag{6}$$

with diffusion coefficient, $D_1$. To obtain survival probability $S(t)$, or the probability that the particle remain in the interval $(-L, L)$ as of time $t$, we first solve equation (6) with absorbing boundary conditions at $x = \pm L$, i.e, $p(\pm L, t) = 0$ using the well-known eigen-function expansion method [30]:

$$p(x,t) = \frac{1}{L} \sum_{k=0}^{\infty} \cos\left[\pi(k+\tfrac{1}{2})\tfrac{x}{L}\right] \exp\left(-\frac{\lambda_k D_1 t}{L^2}\right), \tag{7}$$

where $\lambda_k$ is defined by $\lambda_k = \pi^2 (k+1/2)^2$. Then, the survival probability, $S(t)$, is obtained by integrating equation (7) over $x$ from $-L$ to $L$, that is,

$$S(t) = \int_{-L}^{L} dx\, p(x,t) = \frac{4}{\pi} \sum_{k=0}^{\infty} \frac{(-1)^k}{2k+1} \exp\left(-\frac{\lambda_k D_1 t}{L^2}\right). \tag{8}$$

Substituting equation (8) into equation (5), we obtain the *T*-dependent MFPT, $\langle t \rangle_T$, given by

$$\langle t \rangle_T = \frac{L^2}{D_1} \frac{\frac{4}{\pi} \sum_{k=0}^{\infty} \frac{(-1)^k}{2k+1} \frac{1 - e^{-\lambda_k D_1 T/L^2}(1 + \lambda_k D_1 T/L^2)}{\lambda_k}}{1 - \frac{4}{\pi} \sum_{k=0}^{\infty} \frac{(-1)^k}{2k+1} e^{-\lambda_k D_1 T/L^2}}. \tag{9}$$

At values of $D_1 T/L^2$ larger than about 0.2, equation (9) can be well represented by

$$\langle t \rangle_T \cong \frac{L^2}{D_1} \left[ \frac{1}{2} + \left( \frac{1}{2} - \frac{4 + D_1 T/L^2}{\pi^2} \right) \frac{\frac{4}{\pi} e^{-\pi^2 D_1 T/4L^2}}{1 - \frac{4}{\pi} e^{-\pi^2 D_1 T/4L^2}} \right], \tag{10}$$

which approaches the well-known result for the MFPT, $\langle t \rangle_{T \to \infty} = L^2/2D_1$ [31, 32], in the large-($D_1 T/L^2$) limit [see Appendix A for the derivation of equation (10)]. On the right-hand side of equation (10), $4e^{-\pi^2 D_1 T/4L^2}/\pi$ results from the long-time exponential relaxation of survival probability $S(t)$ given in equation (8). On the other hand, at small $D_1 T/L^2$, equation (9) behaves as

$$\frac{\langle t \rangle_T}{T} = 1 - z' + \frac{5}{2} z'^2 - \frac{37}{4} z'^3 + \mathcal{O}(z'^4), \tag{11}$$

where $z'$ is the dimensionless variable defined by $z' = 4 D_1 T/L^2$ [see Appendix B for the derivation of equation (11)].

## 2.2. Fractional diffusion equation

When a particle's motion is governed by the FDE or equation (2), the Laplace transform, $\hat{p}(x,s)\left[=\int_0^\infty dt e^{-st} p(x,t)\right]$, of $p(x,t)$ can be obtained as [33]:

$$\hat{p}(x,s) = \frac{1}{L}\sum_{k=0}^{\infty} \cos\left[\pi(k+\tfrac{1}{2})\tfrac{x}{L}\right] \frac{1}{s + \lambda_k D_\alpha s^{1-\alpha}/L^2}. \tag{12}$$

This result can be easily obtained simply by replacing $D_1$ with $D_\alpha s^{1-\alpha}$ in the Laplace transform of equation (7), because the Laplace transform of equation (6) with $D_1$ being replaced by $D_\alpha s^{1-\alpha}$ in the Laplace domain is exactly the same as the Laplace transform of equation (2). The inverse Laplace transform of equation (12) yields

$$p(x,t) = \frac{1}{L}\sum_{k=0}^{\infty} \cos\left[\pi(k+\tfrac{1}{2})\tfrac{x}{L}\right] E_\alpha\left(-\frac{\lambda_k D_\alpha t^\alpha}{L^2}\right), \tag{13}$$

where $E_\alpha(t)$ is the Mittag-Leffler function defined by $E_\alpha(t) = \sum_{n=0}^{\infty} t^n/\Gamma(\alpha n + 1)$ [34]. Integrating equation (13) over $x$, we obtain [33]

$$S(t) = \frac{4}{\pi}\sum_{k=0}^{\infty} \frac{(-1)^k}{2k+1} E_\alpha\left(-\frac{\lambda_k D_\alpha t^\alpha}{L^2}\right). \tag{14}$$

Substituting equation (14) into equation (5), we have the *T*-dependent MFPT scaled by *T*, which is given by

$$\frac{\langle t \rangle_T}{T} = \frac{\frac{4}{\pi}\sum_{k=0}^{\infty}\frac{(-1)^k}{2k+1}\left[E_{\alpha,2}\left(-\frac{\lambda_k D_\alpha t^\alpha}{L^2}\right) - E_\alpha\left(-\frac{\lambda_k D_\alpha t^\alpha}{L^2}\right)\right]}{1 - \frac{4}{\pi}\sum_{k=0}^{\infty}\frac{(-1)^k}{2k+1}E_\alpha\left(-\frac{\lambda_k D_\alpha t^\alpha}{L^2}\right)}, \tag{15}$$

where $E_{\alpha,\beta}(t)$ denotes the generalized Mittag-Leffler function defined by $E_{\alpha,\beta}(t) = \sum_{n=0}^{\infty} t^n/\Gamma(\alpha n + \beta)$ [34] and $E_\alpha(t) = E_{\alpha,\beta=1}(t)$. Noting $E_{1,2}(-t) = (1-e^{-t})/t$, we recover equation (9) from equation (15) when $\alpha = 1$.

The MFPT of fractional diffusion motion has a simple power-law dependence on the observation time at long times. Using the long-time asymptotic expansion of $E_{\alpha,\beta}(t)$ for large $|t|$, [34]

$$E_{\alpha,\beta}(t) = -\sum_{n=1}^{\infty}\frac{1}{\Gamma(\beta-\alpha n)}\frac{1}{t^n}, \tag{16}$$

one can show that equation (15) behaves as

$$\begin{aligned}\frac{\langle t \rangle_T}{T} =& \frac{\alpha}{\Gamma(2-\alpha)}\frac{1}{z} - \left[\frac{\alpha R_2}{\Gamma(2-2\alpha)} - \frac{\alpha}{(1-\alpha)\Gamma(1-\alpha)^2}\right]\frac{1}{z^2} \\ &+ \left[\frac{\alpha R_3}{2\Gamma(2-3\alpha)} - \frac{\alpha(3-4\alpha)R_2}{\Gamma(3-2\alpha)\Gamma(1-\alpha)} + \frac{\alpha}{(1-\alpha)\Gamma(1-\alpha)^3}\right]\frac{1}{z^3} \\ &- \left[\frac{\alpha R_4}{6\Gamma(2-4\alpha)} - \frac{\alpha(2-3\alpha)R_3}{3\Gamma(2-3\alpha)\Gamma(2-\alpha)} - \frac{\alpha(1-2\alpha)R_2^2}{2\Gamma(2-2\alpha)^2} \right. \\ &\left. + \frac{\alpha(4-6\alpha)R_2}{\Gamma(3-2\alpha)\Gamma(1-\alpha)^2} - \frac{\alpha}{(1-\alpha)\Gamma(1-\alpha)^4}\right]\frac{1}{z^4} + \mathcal{O}(z^{-5}),\end{aligned} \tag{17}$$

as $D_\alpha T^\alpha/L^2$ approaches infinity. In equation (17), $z$ and $R_n$ respectively denote the dimensionless variable defined by $z = D_\alpha T^\alpha/D_1\langle t_0 \rangle$ and the ratio of the $n$th-order moment, $\langle t_0^n \rangle$, to the $n$th power of the first-order moment, $\langle t_0 \rangle$, i.e., $R_n = \langle t_0^n \rangle / \langle t_0 \rangle^n$, where $t_0$ indicates a first passage time of Brownian motion. By definition, $R_{n=1}$ is simply given by unity. For the current one-dimensional system, the analytical expression of $\langle t_0^n \rangle \left[ = -\int_0^\infty dt_0 t_0^n \, \partial S(t_0)/\partial t_0 \right]$ is obtained as

$$\langle t_0^n \rangle = \frac{2^{2n+2} n!}{\pi^{2n+1}} \left( \frac{L^2}{D_1} \right)^n \sum_{k=0}^{\infty} \frac{(-1)^k}{(2k+1)^{2n+1}} \tag{18}$$

from equation (8). When $n$ is equal to 1, equation (18) reduces to $\langle t_0 \rangle = L^2/2D_1$. In this case, $z$ is explicitly given by $z = D_\alpha T^\alpha / D_1 \langle t_0 \rangle = 2 D_\alpha T^\alpha / L^2$. Then, from the leading term of Eq. (17), we obtain the known asymptotic power-law relation, $\langle t \rangle_T \sim [\alpha/\Gamma(2-\alpha)] L^2 T^{1-\alpha}/2 D_\alpha$, between the MFPT and the observation time for fractional diffusion motion [27]. Equation (17) is one of our main results, which is valid as long as the values of $\alpha$ are less than unity.

It is remarkable that equation (17) is applicable not only to the one-dimensional system but also to a multi-dimensional system with a finite spatial domain. For the latter system, the general long-time asymptotic expansion of the first passage time distribution for the FDE model is given in ref. [35]; the corresponding survival probability has the following long-time asymptotic expression:

$$\begin{aligned} S(T) &= \sum_{n=1}^{\infty} \frac{(-1)^{n+1}}{n!} \frac{\sin(\pi \alpha n) \Gamma(\alpha n)}{\pi} \left( \frac{D_1}{D_\alpha} \right)^n \frac{\langle t_0^n \rangle}{T^{\alpha n}} \\ &= \sum_{n=1}^{\infty} \frac{(-1)^{n+1}}{n! \Gamma(1-\alpha n)} \frac{R_n}{z^n}, \end{aligned} \tag{19}$$

where the reflection formula, $\sin(\pi x)\Gamma(x)/\pi = 1/\Gamma(1-x)$, is used in the second equality. Substituting equation (19) into equation (5), we recover equation (17) in the large-$z$ regime.

On the other hand, in the opposite limit where $D_\alpha T^\alpha/L^2$ approaches zero, equation (15) can be expanded as:

$$\frac{\langle t \rangle_T}{T} = 1 - z' + \left(1 + \frac{3\beta}{2}\right)z'^2 - \left(1 + 4\beta + \frac{4 + 97\alpha + \alpha^2}{24\alpha}\beta^2\right)z'^3 + \mathcal{O}(z'^4), \tag{20}$$

where $\beta = \alpha/(2-\alpha)$ and $z'$ is the dimensionless variable defined by $z' = (4D_\alpha T^\alpha/\alpha^2 L^2)^{\beta/\alpha}$ [see Appendix B for the derivation of equation (20)]. $z'$ is related to $z$ as $z' = (2z/\alpha^2)^{\beta/\alpha}$. Note that equation (20) reduces to equation (11) when $\alpha = 1$.

### 2.3. Scaled Brownian motion

Scaled Brownian motion (SBM) is another model of anomalous diffusion. The position $x(t)$ of particles undergoing SBM obeys the following Langevin equation [36-38]:

$$\frac{\partial}{\partial t}x(t) = \sqrt{2D(t)}\xi(t), \tag{21}$$

where $D(t)$ denotes the time-dependent diffusion coefficient defined by $D(t) = \alpha D_\alpha t^{\alpha-1}/\Gamma(1+\alpha)$. In equation (21), $\xi(t)$ is white Gaussian noise with zero mean and delta function correlation, $\langle \xi(t_1)\xi(t_2) \rangle = \delta(t_1 - t_2)$. In the absence of boundaries, the mean square displacement calculated from equation (21) is the same as equation (1). The Fokker-Planck equation corresponding to equation (21) is given by [36]

$$\frac{\partial}{\partial t} p(x,t) = D(t) \frac{\partial^2}{\partial x^2} p(x,t). \tag{22}$$

By introducing a new time variable, $\tau = \int_0^t dt' D(t')/D_1 = D_\alpha t^\alpha / D_1 \Gamma(1+\alpha)$, equation (22) can be transformed into the classical diffusion equation, equation (6), explicitly, $\partial \tilde{p}(x,\tau)/\partial \tau = D_1 \partial^2 \tilde{p}(x,\tau)/\partial x^2$ with $\tilde{p}(x,\tau) = p(x,t)$. Therefore, the solution of equation (22) under absorbing boundary conditions at $x = \pm L$ can be easily obtained by replacing $t$ with $\tau$ in equation (7):

$$p(x,t) = \frac{1}{L} \sum_{k=0}^{\infty} \cos\left[\pi(k+\tfrac{1}{2})\tfrac{x}{L}\right] \exp\left[-\frac{\lambda_k D_\alpha t^\alpha}{\Gamma(1+\alpha) L^2}\right]. \tag{23}$$

The survival probability for the SBM is then obtained from equation (23) as

$$S(t) = \frac{4}{\pi} \sum_{k=0}^{\infty} \frac{(-1)^k}{2k+1} \exp\left[-\frac{\lambda_k D_\alpha t^\alpha}{\Gamma(1+\alpha) L^2}\right]. \tag{24}$$

The $T$-dependent MFPT for the SBM can then be calculated as equation (5) with equation (24). A detailed discussion on the observation time dependence of the MFPT is presented in Section 3.

**2.4. Fractional Brownian motion**

Fractional Brownian motion (FBM) is a self-similar, nonstationary Gaussian process with stationary increments [39-45]. The FBM is characterized by the Hurst exponent, $H$, which is related to the time exponent, $\alpha$, of the mean square displacement, equation (1) as $H = \alpha/2$ ($0 < H < 1$). For such motion, the autocorrelation function at two different times is given by

$$\langle x(t_1)x(t_2)\rangle = \frac{D_\alpha}{\Gamma(1+\alpha)}(t_1^\alpha + t_2^\alpha - |t_1 - t_2|^\alpha), \tag{25}$$

which reduces to equation (1) when $t_1 = t_2$. The FBM can describe both subdiffusion ($0 < H < 1/2$) and superdiffusion ($1/2 < H < 1$). In the current work, we only consider the case of subdiffusion.

Despite the FBM satisfying the same Fokker-Planck equation, equation (22), as the SBM, the survival probability for the FBM is not given by equation (22). The FBM is a highly non-Markovian process as implied by equation (25), which depends on both $t_1$ and $t_2$. In contrast, the autocorrelation function for the SBM, which is Markovian, depends only on the earlier time, explicitly, $\langle x(t_1)x(t_2)\rangle = 2D_\alpha \min(t_1,t_2)^\alpha / \Gamma(1+\alpha)$ [36, 38]. For the first passage problem of a non-Markovian process, a number of multi-time joint probability distributions characterizing the non-Markovian process are involved in the first passage time distribution or the associated survival probability as explicitly shown in [46, 47]. The Fokker-Planck equation determines only a two-time $(t, t')$ joint or conditional probability distribution that is insufficient for determining the first passage time distribution of a non-Markovian process.

We employ stochastic simulation to obtain the first passage time distribution or the survival probability of FBM, because their analytic expressions are yet to be obtained. Only the long-time asymptotic form of the first passage time distribution is known as $f(t) \sim t^{-(2-H)}$ for a one-dimensional semi-infinite system with a single absorbing boundary [48, 49]. In order to simulate FBM, we used the exact algorithm proposed by Davies and Harte [50]. Our simulation correctly reproduces the mean square displacement, equation (1) with $\alpha = 2H$, in free space and the long-time tail of the first passage time distribution, $f(t) \sim t^{-(2-H)}$, for the semi-infinite system mentioned above.

For a given value of $L$, once the time profile of the survival probability for the FBM is obtained from the simulation, $T$-dependent MFPTs at different values of $T$ for the FBM can be easily calculated by using equation (5) and the time profile of the survival probability. To obtain accurate survival probability of the particle undergoing FBM, the maximum simulation time should be sufficiently larger than the mean first passage time. When observation time $T$ is far shorter than $(L^2/D_\alpha)^{1/\alpha}$, our tracer particle does not reach one of the two absorbing boundaries in the majority of the trajectories. In this case, a direct calculation of the $T$-dependent MFPT from the simulation is inefficient. Equation (5) in the current work provides an efficient way to calculate the $T$-dependent MFPT from the survival probability, which is accurate even when $T \ll (L^2/D_\alpha)^{1/\alpha}$.

**2.5. Wilemski-Fixman approximation**

In this section, let us consider the Wilemski-Fixman (WF) approximation of the survival probability for both the SBM and the FBM, which have the common Fokker-Planck equation, equation (22) [46, 51, 52]. For a particle moving in a semi-infinite interval, $(-\infty, L]$, the distribution, $f(t|x_0)$, of times taken for the particle being initially located at $x_0$ to reach $L$ for the first time can be defined by the following integral equation within the WF approximation [30, 46]:

$$G_0(L,t|x_0) = \int_0^t dt' G_0(L,t-t'|L) f(t'|x_0), \tag{26}$$

where $G_0(x,t|x_0)$ denotes the conditional probability distribution that the particle is found at $x$ at time $t$, given that the particle's initial position was $x_0$. $G_0(x,t|x_0)$ satisfies the normalization condition, i.e., $\int_{-\infty}^{L} dx G_0(x,t|x_0) = 1$ and the reflecting boundary condition at $x = L$, i.e., $\partial G_0(x,t|x_0)/\partial x|_{x=L} = 0$. Equation (26) is exact for a stationary Markov process, but is only approximate for non-stationary or non-Markov processes such as SBM and FBM. Equation (26) is used to find the approximate first passage time distribution for the FBM in ref. [53]. We extend equation (26) to obtain the first passage time distribution of a particle moving in the confined domain, $[-L, L]$ as follows:

$$G_0(\pm L, t | x_0) = \int_0^t dt' \left[ G_0(\pm L, t-t' | L) f_+(t' | x_0) + G_0(\pm L, t-t' | -L) f_-(t' | x_0) \right] \quad (27)$$

Here $f_\pm(t|x_0)$ denote the distributions of times taken to reach $\pm L$ for the first time [see Appendix C for the derivation of equation (27)]. Note that $f_\pm(t|x_0)$ are unnormalized distributions while their sum, $f_+(t|x_0) + f_-(t|x_0) [\equiv f(t|x_0)]$, is normalized. The whole-time integrations of $f_\pm(t|x_0)$ indicate the splitting probabilities that a first-passage event occurs at $\pm L$ between two ends of the domain. In equation (27), $G_0(x,t|x_0)$ is normalized over the confined domain and satisfies the reflecting boundary conditions at both ends of the domain.

In the current case without any external force exerting on a particle and with the choice of $x_0 = 0$, $f_+(t|0)$ is the same as $f_-(t|0)$, i.e., $f_+(t|0) = f_-(t|0) = f(t|0)/2 = f(t)/2$. Equation (27) with $x_0 = 0$ then reduces to

$$G_0(L,t\mid 0) = \frac{1}{2}\int_0^t dt'\left[G_0(L,t-t'\mid L)+G_0(L,t-t'\mid -L)\right]f(t'), \tag{28}$$

From equation (28), we can easily obtain the Laplace domain expressions of the first passage time distribution and the corresponding survival probability, that is,

$$\hat{f}(s) = \frac{2\hat{G}_0(L,s\mid 0)}{\hat{G}_0(L,s\mid L)+\hat{G}_0(L,s\mid -L)} \tag{29a}$$

$$\hat{S}(s) = s^{-1}\left[1-\hat{f}(s)\right]. \tag{29b}$$

On the right-hand side of equation (29a), $\hat{G}_0(x,s\mid x_0)$ is the Laplace transform of the conditional probability distribution or the Green's function, $G_0(x,t\mid x_0)$, which is a solution of equation (22), the Fokker Planck equation governing SBM and FBM, under the initial condition $G_0(x,0\mid x_0)=\delta(x-x_0)$ and the reflecting boundary conditions at $x=\pm L$. Using the similar method to derive equation (23) from equation (7), we can easily obtain the explicit expression of $G_0(x,t\mid x_0)$,

$$G_0(x,t\mid x_0) = \frac{1}{2L} + \frac{1}{L}\sum_{n=1}^{\infty}\cos\left[\frac{n\pi(x+L)}{2L}\right]\cos\left[\frac{n\pi(x_0+L)}{2L}\right]\exp\left[-\left(\frac{n\pi}{2L}\right)^2\frac{D_\alpha t^\alpha}{\Gamma(1+\alpha)}\right], \tag{30}$$

from the Green's function for normal Brownian motion. When the value of $\alpha$ is unity, equation (29) with equation (30) reduces to $\hat{S}(s)=s^{-1}\left[1-\operatorname{sech}\sqrt{sL^2/D_1}\right]$, the Laplace transform of equation (8), which is the exact result obtained for simple Brownian motion. This results because Equation (27) or WF approximation is exact for a stationary Markov process such as Brownian motion, while it is only approximate for non-stationary or non-Markov processes. One can calculate the *T*-dependent MFPT of SBM or FBM with the WF approximation by using equation (5) and the numerical inverse-Laplace transform of equation (29b), which is calculated by the Stehfest algorithm [54] in this work.

## 3. Results and discussion

The observation time dependence of the mean first passage time can serve as a new experimental measure, far more sensitive to the nature of the stochastic process in question than the mean square displacement. Note that, even though subdiffusion models considered in this work yield the same mean square displacement (MSD), whose time profiles all obey equation (1), the observation time dependencies of the MFPT of each model show their own travel length, $L$, or observation time, $T$, dependence (Figures 1 and 2).

The $T$-dependent MFPT linearly increases with the observation time, $T$, when $T$ is far smaller than $\tau \left[ = (L^2 D_\alpha^{-1})^{1/\alpha} \right]$ for all transport models investigated in the current work, as shown in Figure 1. Because only first passage times shorter than $T$ contribute to the $T$-dependent MFPT, the observation time serves as a low-pass filter for first passage times. In the small-$T$ limit, the $T$-dependent MFPT gets close to $T$ for all transport models (see Appendix B).

When $T$ is far greater than $\tau$, the $T$-dependent MFPT of each model shows its own characteristic behavior. For SBM, FBM and their WF approximation, the $T$-dependent MFPT, $\langle t \rangle_T / \tau$, saturates to a finite value in the large-$T$ limit, whereas for the FDE model, the value of $\langle t \rangle_T / \tau$ keeps increasing with observation time $T$, obeying the following power-law dependence, $\langle t \rangle_T / \tau \sim T^{1-\alpha}$ on the observation time. The MFPT in the large-$T$ limit is the same as $\int_0^\infty dt S(t)$; the finite MFPT of SBM and FBM results because their survival probabilities $S(t)$ decay to zero following a stretched exponential and a simple exponential functions, respectively [55-57] (Figure 3A and inset). For SBM and its WF approximation, analytical expressions of the MFPT in the large *T* limit can be obtained simply by integrating equation (24) over the entire time region and taking the small-*s* limit value of equation (29b) with equation (30), respectively:

$$\langle t \rangle_{T\to\infty}^{\text{SBM}} = \frac{\zeta(1+\tfrac{2}{\alpha},\tfrac{1}{4}) - \zeta(1+\tfrac{2}{\alpha},\tfrac{3}{4})}{\pi}\Gamma(1+\tfrac{1}{\alpha})\left(\frac{\Gamma(1+\alpha)L^2}{4\pi^2 D_\alpha}\right)^{1/\alpha}, \qquad (31a)$$

$$\langle t \rangle_{T\to\infty}^{\text{WF}} = 4(4^{1/\alpha}-1)\zeta(\tfrac{2}{\alpha})\Gamma(1+\tfrac{1}{\alpha})\left(\frac{\Gamma(1+\alpha)L^2}{4\pi^2 D_\alpha}\right)^{1/\alpha}, \qquad (31b)$$

where $\zeta(s,a)$ denotes the generalized Riemann zeta function defined by $\zeta(s,a) = \sum_{n=0}^{\infty}(n+a)^{-s}$ and $\zeta(s) = \zeta(s, a=0)$. For the FBM, the analytical expression of the MFPT is unavailable but it can be shown that $\langle t \rangle_{T\to\infty}^{\text{FBM}}$ has the same *L*-dependence as equations (31a) and (31b), i.e., $\langle t \rangle_{T\to\infty}^{\text{FBM}} \sim L^{2/\alpha}$ [58, 59] (Figure 3B). On the other hand, the survival probability for the FDE model decays with an algebraic tail at long times, i.e., $S(t) \sim t^{-\alpha}$ $(0 < \alpha < 1)$. The integration, $\int_0^T dt S(t)$, of the survival probability over time from 0 to *T* is then dominated by the long-time behavior of $S(t)$ for large *T*. As a result, $\int_0^T dt S(t)$ behaves as $\int_0^T dt S(t) \sim \int_0^T dt\, t^{-\alpha} \sim T^{1-\alpha}$ at large *T*. This can also be understood from equation (17) in the large-*T* or large-*z* limit, where the leading-order term is given by $\langle t \rangle_T / T \sim [\alpha/\Gamma(2-\alpha)]z^{-1}$, or $\langle t \rangle_T \sim [\alpha/\Gamma(2-\alpha)]L^2 T^{1-\alpha}/2D_\alpha$. The higher-order terms in equation (17) describe how the *T*-dependent MFPT for the FDE model deviates from asymptotic behavior as $\tilde{T}$ or $z(=2\tilde{T}^\alpha)$ increases. When the value of $\alpha$ is given by 0.01 (1), about 65 (90) % of all first passage events have occurred as of $\tilde{T}=1$, and the relative error of equation (17) keeping up to the fourth-order term is less than 10% for any value of $\alpha$ less than unity.

In Figure 2, the *T*-dependent MFPT of the FDE model quadratically depends on *L* in the small-*L* limit, which resembles the diffusive scaling of the MFPT, $\langle t_0 \rangle (= L^2/2D_1)$, for normal Brownian motion. The $L^2$-scaling behavior results from the leading-order term on the right hand side of equation (17), dominant in the small-*L* limit, where equation (17) reduces to $\langle t \rangle_T / T$ $\sim [\alpha/\Gamma(2-\alpha)]z^{-1} = [\alpha/\Gamma(2-\alpha)]D_1\langle t_0 \rangle / D_\alpha T^\alpha = [\alpha/\Gamma(2-\alpha)]L^2/2D_\alpha T^\alpha$ $(L^2/2DT^\alpha \ll 1)$.

The SBM, FBM, and their WF approximation version show a different *L*-dependence, $\langle t \rangle_T / T \sim L^{2/\alpha}$, from the FDE model at small values of $L/(D_\alpha T^\alpha)^{1/2}$, which is shown in equation (31) and Figures 2A-2D, and 3B. As shown shortly, $\langle t \rangle_T / T$ is a function of $L/(D_\alpha T^\alpha)^{1/2}$ only, so that the small *L* limit value of $\langle t \rangle_T / T$ is the same as its large *T* limit value for these transport models, as long as it exists. Accordingly, the MFPTs of the SBM, FBM, and their WF approximation version depend on *L* as $\langle t \rangle_{T\to\infty} \sim L^{2/\alpha}$.

As *L* increases, for every transport model investigated, $\langle t \rangle_T / T$ deviates from the small *L*-scaling behavior and monotonically increases, approaching unity. When $L/(D_\alpha T^\alpha)^{1/2}$ is much greater than unity, most first passage events occurs at times greater than observation time *T*. That is to say, the relative contribution of first passage events occurring at times less than *T* becomes smaller, as *L* increases. In the large-*L* limit, the mean of the first passage times less than *T* approaches *T* for any transport model (Appendix B).

The observation time dependence of the MFPT is closely related to the travel length dependence of the MFPT. This is because the survival probability, $S(T)$, is essentially a function of the dimensionless variable, $\tilde{T}(\equiv T/\tau)$, with $\tau$ being defined as $(L^2 D_\alpha^{-1})^{1/\alpha}$, i.e., $S(T) \equiv g(\tilde{T})$ for every transport model investigated in this work [see equations (8), (14), and (24) for examples]. Consequently, the $T$-dependent MFPT scaled by $T$ is also a function of $\tilde{T}$, that is, $\langle t \rangle_T / T \equiv h(\tilde{T})$, which can be clearly seen by following rearrangement of equation (5):

$$\frac{\langle t \rangle_T}{T} = \frac{T^{-1} \int_0^T dt\, S(t) - S(T)}{1 - S(T)}$$
$$= \frac{\tilde{T}^{-1} \int_0^{\tilde{T}} d\tilde{t}\, g(\tilde{t}) - g(\tilde{T})}{1 - g(\tilde{T})}, \tag{32}$$

with $\tilde{t}$ denoting $t/\tau$. Equation (32) implies that, so long as the value of $\tilde{T}[= T/(L^2 D_\alpha^{-1})^{1/\alpha}]$ is kept constant for a given $\alpha$, the value of $\langle t \rangle_T / T$ is invariant under any change in the values of $T$ and $L$ (see Appendix D).

The difference between the MFPT values of the SBM, FBM, and their WF approximation for the same values of $L$, $\alpha$, and $D_\alpha$ can be understood in terms of the behavior of the normalized displacement distribution defined by $p(x,t)/S(t)$ near boundaries, $x = \pm L$. In the long-time limit, $p(x,t)/S(t)$ approaches a stationary distribution, $\lim_{t \to \infty} p(x,t)/S(t) \equiv p_\infty(x)$ [57, 59]. The analytical expressions of $p_\infty(x)$ for the FDE model and the SBM are given below:

$$p_\infty(x) = \frac{2}{\pi^2 L} \sum_{k=0}^{\infty} \frac{\cos\left[\pi(k+\tfrac{1}{2})\tfrac{x}{L}\right]}{(k+\tfrac{1}{2})^2} = \frac{1 - |x/L|}{L}, \quad \text{(FDE)} \tag{33a}$$

$$p_\infty(x) = \frac{\pi}{4L} \cos\left(\frac{\pi x}{2L}\right). \quad \text{(SBM)} \tag{33b}$$

Equations (33a) and (33b) can be obtained by considering only the leading-order term in the long-time asymptotic expansion of the Mittag-Leffler function in equations (13) and (14), and only the leading-order terms of equations (23) and (24) at long times, respectively. Using the orthogonality relation, $\frac{1}{L}\int_{-L}^{L} dx \cos\left[\pi(k+\frac{1}{2})\frac{x}{L}\right]\cos\left[\pi(l+\frac{1}{2})\frac{x}{L}\right] = \delta_{kl}$, it can be shown that the second equality in equation (33a) holds. For the FBM and its WF approximation, it is difficult to derive the analytical expressions of $p_\infty(x)$, but their spatial profiles can be obtained from stochastic simulation of the FBM and from the numerical inverse Laplace transform of equation (C.11) with equations (C.12) and (30). At either absorbing boundary, $p_\infty(x)$ behaves as $p_\infty(x) \sim \tilde{x}^\phi$ for every model investigated, where $\tilde{x}$ designates $\tilde{x} = 1 - |x/L|$ (Figure 3D) [57]. It is known that $\phi$ is related to the persistence exponent, $\theta$, as $\phi = 2\theta/\alpha$, where $\theta$ characterizes a long-time tail, $\rho(t) \sim t^{-1-\theta}$, of the persistence time distribution, $\rho(t)$ [60-62]. Here, a persistence time designates a time interval between consecutive zeros of a stochastic process, $x(t)$. A large value of $\theta$ indicates a less persistent process, in other words, a more thorough exploration over space. Thus, a less persistent process with a greater value of $\theta$ searches more efficiently for specific targets, which is absorbing boundaries in our problem, resulting in lower values of $p_\infty(x)$ near the absorbing boundaries (Figure 3C or 3D) and a faster decay of survival probability (Figure 3A). Among the subdiffusion models investigated, the FBM and the SBM results have the highest and the lowest $\phi$ values, respectively, and the WF result is in-between, implying that a stationary Markovian process, assumed in the WF approximation, is more advantageous than a nonstationary Markovian process, that is, SBM with respect to searching for specific targets, in the case of subdiffusion [63].

As shown in Figure 3D, for the FBM with $\alpha = 0.3$, the numerical value of $\phi$ is found to be 5.67, which is in good agreement with the $\alpha$-dependence of $\phi$, i.e., $\phi = (2-\alpha)/\alpha$, previously reported in [57]. For both the FDE model and the SBM, the value of $\phi$ is found to be unity, regardless of $\alpha$; $p_\infty(x)$ given in equations (33a) or (33b) vanishes linearly as $\tilde{x}/L$ or $\pi^2 \tilde{x}/8L$ at small $\tilde{x}$. For the WF approximation version of SBM and FBM, $\phi$ shows an interesting dependence on $\alpha$ (Figures 3E and 3F). There are three phases; (1) $\phi$ resembles the $\alpha$-dependence for the FBM at values of $\alpha$ larger than roughly 0.8, (2) $\phi$ varies linearly at values of $\alpha$ between 0.45 and 0.8, and (3) $\phi$ eventually saturates to two at values of $\alpha$ less than 0.45. This result implies that the WF result might be a good approximation of the FBM at values of $\alpha$ larger than 0.8, where the non-Markovian effect is not yet significant. Note that the SBM is also a Markovian process but the $\alpha$-dependence of $\phi$ for the SBM is qualitatively different from the FBM even at values of $\alpha$ close to unity (Figure 3F).

## 4. Conclusion

For fractional diffusions equations, for which the MFPT does not exist, the observation time, $T$, dependent MFPT defined in equation (4) or (5) is investigated and compared with normal diffusion and three different models of anomalous subdiffusion, namely, scaled Brownian motion, fractional Brownian motion, and WF approximation or the stationary Markov approximation of SBM and FBM. The observation time dependence of the MFPT can serve as a new experimental measure, far more sensitive to the nature of the stochastic process in question than the mean square displacement. When $\tilde{T}\left[=T/(L^2 D_\alpha^{-1})^{1/\alpha}\right]$ is small, the $T$-dependent MFPT is linearly proportional to observation time $T$ for every transport model investigated. As $\tilde{T}\left[=T/(L^2 D_\alpha^{-1})^{1/\alpha}\right]$ increases, the $T$-dependent MFPT of SBM, FBM and their WF approximation approach finite values, which commonly have the power-law dependence, $L^{2/\alpha}$, on travel length $L$ in contrast with the $T$-dependent MFPT of FDE, which diverges in the long time limit, following $T^{1-\alpha}$. The $T$ dependent MFPT of FDE shows the same $L^2$ dependence on travel length $L$ as the normal Brownian motion, in contrast with other models of anomalous diffusion, SBM, FBM, and their WF approximation.

Equations (17) and (20) describe the asymptotic behaviors of the $T$-dependent MFPT of the FDE model in terms of the dimensionless observation time, $\tilde{T}\left[=T/(L^2 D_\alpha^{-1})^{1/\alpha}\right]$, at large and small $\tilde{T}$ regime, respectively. Equation (17) is applicable not only to one-dimensional system but also to a multi-dimensional system in a finite domain. The asymptotic behavior of the $T$-dependent MFPT of the normal diffusion is given in equations (10) and (11) for large and small $\tilde{T}$ values, respectively.

## Acknowledgements

The authors gratefully acknowledge Mr. Luke Bates for his careful reading of our manuscript. This work was supported by the Creative Research Initiative Project program (2015R1A3A2066497) of the National Research Foundation of Korea (NRF) and the NRF grant (MSIT) (2019R1C1C1004576) funded by the Korea government.

**Appendix A. Derivation of equation (10)**

Here, we derive equation (10), which is the large-$\tilde{T}$ expansion of equation (9). Letting $u$ denote $u = e^{-\pi^2 D_1 T/L^2} = e^{-\pi^2 \tilde{T}}$, equation (9) can be rewritten as

$$\frac{D_1 \langle t \rangle_T}{L^2} = \frac{\frac{4}{\pi} \sum_{k=0}^{\infty} \frac{(-1)^k}{(2k+1)\lambda_k} \left[ 1 - u^{(k+\frac{1}{2})^2} \left[ 1 - \ln u^{(k+\frac{1}{2})^2} \right] \right]}{1 - \frac{4}{\pi} \sum_{k=0}^{\infty} \frac{(-1)^k}{2k+1} u^{(k+\frac{1}{2})^2}}. \quad (A.1)$$

Noting that $(4/\pi)\sum_{k=0}^{\infty}(-1)^k/(2k+1)\lambda_k$ on the right-hand side of equation (A.1) is exactly equal to one half, the large-$\tilde{T}$ or small-$u$ expansion of equation (A.1) is then given by

$$\frac{D_1 \langle t \rangle_T}{L^2} = \frac{1}{2} + \left( \frac{1}{2} + \frac{\ln u - 4}{\pi^2} \right) \sum_{k=1}^{\infty} \left( \frac{4u^{1/4}}{\pi} \right)^k. \quad (A.2)$$

When $4u^{1/4}/\pi$ is less than unity, equation (A.2) can be expressed as

$$\frac{D_1 \langle t \rangle_T}{L^2} = \frac{1}{2} + \left( \frac{1}{2} + \frac{\ln u - 4}{\pi^2} \right) \frac{4u^{1/4}/\pi}{1 - 4u^{1/4}/\pi}, \quad (A.3)$$

which is equivalent to equation (10) in the main text. At values of $\tilde{T}$ larger than about 0.2, equation (9) can be well represented by equation (A.3) or (10).

## Appendix B. Derivation of equations (11) and (20)

In this appendix, we derive equation (20), which is the small-$\tilde{T}$ expansion of equation (26) with $g(\tilde{T})$, or equivalently $S(T)$ being given by equation (12). Equation (11) is a special case of equation (20) at $\alpha=1$. First, let us consider the Laplace transform of equation (12), explicitly, $g(\tilde{t})$ with respect to $\tilde{t}\left[=t/\tau=t/(L^2 D_\alpha^{-1})^{1/\alpha}\right]$, which can be obtained as

$$\hat{g}(\tilde{s})=\frac{1-\operatorname{sech}(\tilde{s}^{\alpha/2})}{\tilde{s}}, \tag{B.1}$$

where $\tilde{s}$ denotes the dimensionless Laplace variable defined by $\tilde{s}=s\tau$. In the large-$\tilde{s}$ limit, which corresponds to the small-$\tilde{t}$ limit in the time domain, equation (B.1) reduces to

$$\hat{g}(\tilde{s})\cong\frac{1-2e^{-\tilde{s}^{\alpha/2}}}{\tilde{s}}=\frac{1}{\tilde{s}}-\frac{2}{\tilde{s}}\sum_{n=0}^{\infty}\frac{(-1)^n \tilde{s}^{\alpha n/2}}{n!}, \tag{B.2}$$

the inverse Laplace transformation of which gives

$$g(\tilde{t})\cong 1-2W(-\tilde{t}^{-\alpha/2};-\tfrac{\alpha}{2},1), \tag{B.3}$$

where $W(x;a,b)$ denotes the Wright function defined by [34]

$$W(x;a,b)=\sum_{n=0}^{\infty}\frac{x^n}{n!\Gamma(an+b)}. \tag{B.4}$$

When $\alpha=1$, equation (B.3) can be written as

$$g(\tilde{t})\cong 1-2\operatorname{erfc}(\tilde{t}^{-1/2}/2), \tag{B.5}$$

where erfc(x) denotes the complementary error function defined by $\text{erfc}(x) = 1 - \frac{2}{\sqrt{\pi}} \int_0^x dt\, e^{-t^2}$.

Substituting equation (B.3) into equation (26), we have

$$\frac{\langle t \rangle_T}{T} \cong 1 - \frac{\tilde{T}^{-1} \int_0^{\tilde{T}} d\tilde{t}\, W(-\tilde{t}^{-\alpha/2}; -\frac{\alpha}{2}, 1)}{W(-\tilde{T}^{-\alpha/2}; -\frac{\alpha}{2}, 1)}. \tag{B.6}$$

In the small-$\tilde{T}$ limit, we need the large-$x$ asymptotic expansion of $W(-x; a, b)$, which is given by [64]

$$W(-x; a, b) = \frac{X^{1/2-b} e^{-X}}{2\pi} \sum_{j=0}^{\infty} (-1)^j A_j X^{-j}, \quad (0 < \sigma < 1) \tag{B.7}$$

where $\sigma$ and $X$ are defined by $\sigma = -a$ and $X = (1-\sigma)(\sigma^\sigma x)^{1/(1-\sigma)}$, respectively. In equation (B.7), $A_0$, $A_1$ and $A_2$ are given by

$$\begin{aligned}
A_0 &= \left(\frac{2\pi}{1-\sigma}\right)^{1/2} \left(\frac{\sigma}{1-\sigma}\right)^{1/2-b}, \\
A_1 &= -\frac{A_0}{24a}\left[(2+a)(1+2a) - 12b(1+a-b)\right], \\
A_2 &= \frac{A_0}{1152a^2}\Big[(2+a)(1+2a)(2-19a+2a^2) + 24b(1+a)(2+7a-6a^2) \\
&\quad - 24b^2(4-5a-20a^2) - 96b^3(1+5a) + 144b^4\Big].
\end{aligned} \tag{B.8}$$

When $a = -1/2$ and $b = 1$, equation (B.7) is just the large-$x$ asymptotic expansion of $\text{erfc}(x/2)$. In a more general case where $a = -\alpha/2$ and $b = 1$, the substitution of equation (B.7) into equation (B.6) yields

$$\frac{\langle t \rangle_T}{T} \cong 1 - \frac{\sum_{j=0}^{\infty}(-1)^j A_j (c/\tilde{T}^\beta)^{-j-1/2} \int_0^1 du\, e^{-c/(\tilde{T}u)^\beta} u^{\beta(j+1/2)}}{\sum_{j=0}^{\infty}(-1)^j A_j (c/\tilde{T}^\beta)^{-j-1/2} e^{-c/\tilde{T}^\beta}}, \qquad (B.9)$$

where $c$ and $\beta$ indicate $c = (1-\alpha/2)(\alpha/2)^\beta$ and $\beta = \alpha/(2-\alpha)$, respectively. In the numerator of the second term on the right-hand side of equation (B.9), the integration over $u$ from zero to unity can be calculated as

$$\int_0^1 du\, e^{-c/(\tilde{T}u)^\beta} u^{\beta(j+1/2)} = \beta^{-1} \left( \frac{c}{\tilde{T}^\beta} \right)^{j+1/2+1/\beta} \Gamma\left( -j - \frac{1}{2} - \frac{1}{\beta}, \frac{c}{\tilde{T}^\beta} \right), \qquad (B.10)$$

where $\Gamma(s,x)$ denotes the upper incomplete gamma function defined by $\Gamma(s,x) = \int_x^\infty dt\, e^{-t} t^{s-1}$. Finally, we have

$$\frac{\langle t \rangle_T}{T} \cong 1 - \frac{\beta^{-1} \sum_{j=0}^{\infty}(-1)^j A_j Y^{1/\beta} \Gamma\left( -j - \frac{1}{2} - \frac{1}{\beta}, Y \right)}{\sum_{j=0}^{\infty}(-1)^j A_j Y^{-j-1/2} e^{-Y}} \qquad (B.11)$$

with $Y$ denoting $Y = (1-\alpha/2)(\alpha/2)^\beta / \tilde{T}^\beta$. The large-$Y$ expansion of equation (B.11) up to the third order in $Y^{-1}$ is just the same as equation (18) in the main text, noting that $\beta Y = (\alpha/2)^{\beta+1}/\tilde{T}^\beta = (\alpha^2/4\tilde{T}^\alpha)^{\beta/\alpha} = 1/z'$.

**Appendix C. Derivation of equation (27)**

Here, we derive equation (27) from a reaction-diffusion equation, which is given by

$$\frac{\partial}{\partial t}G(x,t|x_0) = \left[D(t)\frac{\partial^2}{\partial x^2} - K(x)\right]G(x,t|x_0)$$
$$\equiv \mathcal{L}(t)G(x,t|x_0), \qquad (C.1)$$

where $G(x,t|x_0)$ denotes the Green function, or the conditional probability of finding a particle at position $x$ without any reaction event by time $t$, given that the initial position of the particle was $x_0$. In equation (C.1), $D(t)$ is the time-dependent diffusion coefficient defined in equation (19) and $K(x)$ is the reaction sink function. In our case, the explicit expression of $K(x)$ is given by $K(x) = \kappa[\delta(x-x_+) + \delta(x-x_-)]$, where $\kappa$ denotes the reaction sink strength and $x_{\pm} = \pm L$. The formal solution of equation (C.1) is given by

$$G(x,t|x_0) = T_L \exp\left(\int_0^t dt' \mathcal{L}(t')\right)\delta(x-x_0), \qquad (C.2)$$

where $T_L$ is the left-time-ordering operator that interchanges the time-dependent operators in such a way that time increases from right to left [65]. Using the Dyson decomposition [66], equation (C.2) can be rewritten as

$$G(x,t|x_0) = T_L \exp\left(\int_0^t dt' \mathcal{L}_0(t')\right)\delta(x-x_0)$$
$$- \int_0^t dt_1 T_L \exp\left(\int_{t_1}^t dt' \mathcal{L}_0(t')\right) K(x) T_L \exp\left(\int_0^{t_1} dt' \mathcal{L}(t')\right)\delta(x-x_0), \qquad (C.3)$$

where $\mathcal{L}_0(t)$ stands for $D(t)\partial_x^2$. Inserting $\int dx_1 \delta(x-x_1)$ on the left of $K(x)$ in the second term on the right-hand side of equation (C.3), we have

$$G(x,t|x_0) = G_0(x,t|x_0) - \int_0^t dt_1 \int dx_1 G_0(x,t|x_1,t_1)K(x_1)G(x_1,t_1|x_0), \tag{C.4}$$

where $G_0(x,t|x_1,t_1)$ is the reaction-free Green function defined by $G_0(x,t|x_1,t_1) = T_L \exp\left(\int_{t_1}^t dt' \mathcal{L}_0(t')\right)\delta(x-x_1)$ and $G_0(x,t|x_1,t_1=0) = G_0(x,t|x_1)$. For the current one-dimensional system, $G_0(x,t|x_0)$ is given by equation (24). The iterative solution of the integral equation (C.4) is then given by

$$\begin{aligned}G(x,t|x_0) &= G_0(x,t|x_0) - \int_0^t dt_1 \int dx_1 G_0(x,t|x_1,t_1)K(x_1)G_0(x_1,t_1|x_0) \\ &+ \int_0^t dt_2 \int_0^{t_2} dt_1 \int dx_2 \int dx_1 G_0(x,t|x_2,t_2)K(x_2)G_0(x_2,t_2|x_1,t_1) \\ &\times K(x_1)G_0(x_1,t_1|x_0) - \cdots.\end{aligned} \tag{C.5}$$

Employing the WF approximation in equation (C.5), where the time shifting in the Green function is allowed, i.e., $G(x_i,t_i|x_{i-1},t_{i-1}) = G(x_i,t_i-t_{i-1}|x_{i-1})$, equation (C.5) can be written as

$$\begin{aligned}G(x,t|x_0) &= G_0(x,t|x_0) - \int_0^t dt_1 \int dx_1 G_0(x,t-t_1|x_1)K(x_1)G_0(x_1,t_1|x_0) \\ &+ \int_0^t dt_2 \int_0^{t_2} dt_1 \int dx_2 \int dx_1 G_0(x,t-t_2|x_2)K(x_2)G_0(x_2,t_2-t_1|x_1) \\ &\times K(x_1)G_0(x_1,t_1|x_0) - \cdots.\end{aligned} \tag{C.6}$$

The Laplace transform of equation (C.6) can then be obtained as

$$\begin{aligned}\hat{G}(x,s|x_0) &= \hat{G}_0(x,s|x_0) - \int dx_1 \hat{G}_0(x,s|x_1)K(x_1)\hat{G}_0(x_1,s|x_0) \\ &+ \int dx_2 \int dx_1 \hat{G}_0(x,s|x_2)K(x_2)\hat{G}_0(x_2,s|x_1)K(x_1)\hat{G}_0(x_1,s|x_0) - \cdots.\end{aligned} \tag{C.7}$$

Substituting $K(x) = \kappa[\delta(x-x_+) + \delta(x-x_-)]$ into equation (C.7), we have

$$\hat{G}(x,s|x_0) = \hat{G}_0(x,s|x_0) - \kappa \sum_{i=+,-} \hat{G}_0(x,s|x_i)\hat{G}_0(x_i,s|x_0)$$
$$+ \kappa^2 \sum_{i=+,-} \sum_{j=+,-} \hat{G}_0(x,s|x_i)\hat{G}_0(x_i,s|x_j)\hat{G}_0(x_j,s|x_0) \qquad (C.8)$$
$$- \kappa^3 \sum_{i=+,-} \sum_{j=+,-} \sum_{k=+,-} \hat{G}_0(x,s|x_i)\hat{G}_0(x_i,s|x_j)\hat{G}_0(x_j,s|x_k)\hat{G}_0(x_k,s|x_0) + \cdots.$$

Using the (2×2)-dimensional matrix, **G**, defined by

$$\mathbf{G} = \begin{pmatrix} \hat{G}_0(x_+,s|x_+) & \hat{G}_0(x_+,s|x_-) \\ \hat{G}_0(x_-,s|x_+) & \hat{G}_0(x_-,s|x_-) \end{pmatrix}, \qquad (C.9)$$

Equation (C.8) can be rewritten in a more concise form as

$$\hat{G}(x,s|x_0) = \hat{G}_0(x,s|x_0) - \kappa \sum_{i=+,-} \sum_{j=+,-} \hat{G}_0(x,s|x_i)(\mathbf{I})_{ij}\hat{G}_0(x_i,s|x_0)$$
$$+ \kappa \sum_{i=+,-} \sum_{j=+,-} \hat{G}_0(x,s|x_i)(\kappa\mathbf{G})_{ij}\hat{G}_0(x_j,s|x_0)$$
$$- \kappa \sum_{i=+,-} \sum_{j=+,-} \hat{G}_0(x,s|x_i)(\kappa^2\mathbf{G}^2)_{ij}\hat{G}_0(x_j,s|x_0) + \cdots \qquad (C.10)$$
$$= \hat{G}_0(x,s|x_0) - \kappa \sum_{i=+,-} \sum_{j=+,-} \hat{G}_0(x,s|x_i)(\mathbf{I} - \kappa\mathbf{G} + \kappa^2\mathbf{G}^2 - \cdots)_{ij}\hat{G}_0(x_i,s|x_0)$$
$$= \hat{G}_0(x,s|x_0) - \kappa \sum_{i=+,-} \sum_{j=+,-} \hat{G}_0(x,s|x_i)([\mathbf{I} + \kappa\mathbf{G}]^{-1})_{ij}\hat{G}_0(x_i,s|x_0)$$

where **I** and $(\mathbf{G})_{ij}$ denote the (2×2)-dimensional identity matrix and $\hat{G}_0(x_i,s|x_j)$, respectively. The solution under absorbing boundary conditions at two ends of the box can be obtained by taking the large-$\kappa$ limit of equation (C.10):

$$\hat{G}(x,s|x_0) = \hat{G}_0(x,s|x_0) - \left[\hat{G}_0(x,s|x_+)\hat{f}_+(s|x_0) + \hat{G}_0(x,s|x_-)\hat{f}_-(s|x_0)\right], \qquad (C.11)$$

where $\hat{f}_\pm(s|x_0)$ are defined by

$$\hat{f}_\pm(s|x_0) = \frac{\hat{G}_0(x_\mp,s|x_\mp)\hat{G}_0(x_\pm,s|x_0) - \hat{G}_0(x_\pm,s|x_\mp)\hat{G}_0(x_\mp,s|x_0)}{\det(\mathbf{G})}. \quad (C.12)$$

In equation (C.12), $\det(\mathbf{G})$ stands for the determinant of the matrix $\mathbf{G}$ defined by equation (C.9). In addition, thanks to the symmetry of the current system, the following equalities hold; $\hat{G}_0(x_+,s|x_+) = \hat{G}_0(x_-,s|x_-)$ and $\hat{G}_0(x_+,s|x_-) = \hat{G}_0(x_-,s|x_+)$ as can be directly seen from equation (30). Integrating both sides of equation (C.11) over $x$ from $x_-$ to $x_+$, and noting that $\int_{x_-}^{x_+} dx\, G(x,t|x_0) = S(t|x_0)$, which denotes the survival probability conditioned on the initial position, $x_0$, and $\int_{x_-}^{x_+} dx\, G_0(x,t|x_0) = 1$, or $\int_{x_-}^{x_+} dx\, \hat{G}_0(x,s|x_0) = 1/s$, we have

$$\hat{S}(s|x_0) = \frac{1}{s} - \frac{1}{s}\left[\hat{f}_+(s|x_0) + \hat{f}_-(s|x_0)\right], \quad (C.13)$$

indicating that the square-bracketed factor, $\hat{f}_+(s|x_0) + \hat{f}_-(s|x_0)$, corresponds to the Laplace transform, $\hat{f}(s|x_0)$, of the first passage time distribution conditioned on the initial position, $x_0$. In addition, because $\hat{G}(x,s|x_0)$ in equation (C.11) satisfies the absorbing boundary conditions at $x = x_\pm$, in short, $\hat{G}(x_\pm,s|x_0) = 0$, equation (C.11) at $x = x_\pm$ gives

$$\hat{G}_0(x_\pm,s|x_0) = \hat{G}_0(x_\pm,s|x_+)\hat{f}_+(s|x_0) + \hat{G}_0(x_\pm,s|x_-)\hat{f}_-(s|x_0), \quad (C.14)$$

the inverse Laplace transformation of which is given by

$$G(x_{\pm},t|x_0) = \int_0^t dt' \left[ G_0(x_{\pm},t-t'|x_+)f_+(t'|x_0) + G_0(x_{\pm},t-t'|x_-)f_-(t'|x_0) \right], \tag{C.15}$$

which is equivalent to equation (27) in the main text. Equation (C15) tells us that $f_{\pm}(t|x_0)$ are, in fact, the distributions of times taken to reach $x_{\pm}$ for the first time, given that the particle's initial position was $x_0$. On the right-hand side of equation (C.15), the first (second) term accounts for the contribution that a particle first reached $x_{+(-)}$ at an earlier time, $t'$, and is found at $x_+$ at time $t$, allowing multiple visits to $x_+$ between $t'$ and $t$.

$\hat{f}_{\pm}(0|x_0) \left[ = \int_0^\infty dt\, f_{\pm}(t|x_0) \right]$ indicate the splitting probabilities that a first-passage event occurs at $x_{\pm}$ between two ends of the domain. From the small-$s$ expansion of equation (C.13) with equations (C.12) and (30), the explicit expressions of $\hat{f}_{\pm}(0|x_0)$ can be obtained as

$$\hat{f}_{\pm}(0|x_0) = \frac{1}{2} \mp \frac{1}{2(1-4^{-1/\alpha})\zeta(2/\alpha)} \sum_{k=0}^{\infty} \frac{\cos\left[(2k+1)\pi(x_0+L)/2L\right]}{(2k+1)^{2/\alpha}}. \tag{C.16}$$

with $\zeta(z)$ denoting the Riemann zeta function defined by $\zeta(z) = \sum_{k=0}^{\infty} k^{-z}$. When $2/\alpha$ is an even integer, i.e., $2/\alpha = 2n$, equation (C.16) can be rewritten as [67]

$$\hat{f}_{\pm}(0|x_0) = \frac{1}{2} \pm \frac{(-1)^{n+1}(2\pi)^{2n} E_{2n-1}(\frac{x_0+L}{2L})}{8(4^n-1)\Gamma(2n)\zeta(2n)}, \tag{C.17}$$

where $E_q(x)$ denotes the $q$th-order Euler polynomial defined by

$$E_q(x) = \sum_{k=0}^{q} \frac{1}{2^k} \sum_{m=0}^{k} \frac{(-1)^m k!}{m!(k-m)!}(x+m)^q. \tag{C.18}$$

When $\alpha=1$, or equivalently $n=1$ in equation (C.17), equation (C.17) yields the simple, well-known results for the Brownian motion, which are given by $\hat{f}_\pm(0|x_0)=1/2\pm x_0/2L$ [30].

**Appendix D. Relationship between observation time dependence and travel length dependence of the *T*-dependent MFPT**

The theoretical and simulation results of the *T*-dependence of $\langle t \rangle_T / (L^2 D_\alpha^{-1})^{1/\alpha}$ shown in Figure 1 can be easily converted into the results of the *L*-dependence of $\langle t \rangle_T / T$ in Figure 2. Once a *n*-point data set, $\{(\tilde{T}_i = T_i / (L^2 D_\alpha^{-1})^{1/\alpha}, h(\tilde{T}_i)) | 1 \leq i \leq n\}$, is made using equation (5) or (32) with the time profile of $S(t)$ obtained for a given *L*, the same data set can be obtained by keeping *T* constant but varying *L*, explicitly, $\{(\tilde{T}_i = T / (L_i^2 D_\alpha^{-1})^{1/\alpha}, h(\tilde{T}_i)) | 1 \leq i \leq n\}$. Using this fact, a data set, $\{(\tilde{T}_i, h(\tilde{T}_i)\tilde{T}_i) | 1 \leq i \leq n\}$, for the *T*-dependence of $\langle t \rangle_T / \tau [= \langle t \rangle_T \tilde{T} / T = h(\tilde{T})\tilde{T}]$ can be converted into the data set, $\{\tilde{T}_i^{-\alpha/2}, h(\tilde{T}_i)) | 1 \leq i \leq n\}$, for the *L*-dependence of $\langle t \rangle_T / T [= h(\tilde{T})]$. Here, $\tilde{T}_i^{-\alpha/2}$ corresponds to the dimensionless length scale, i.e., $\tilde{T}_i^{-\alpha/2} = L_i / (D_\alpha T^\alpha)^{1/2}$.

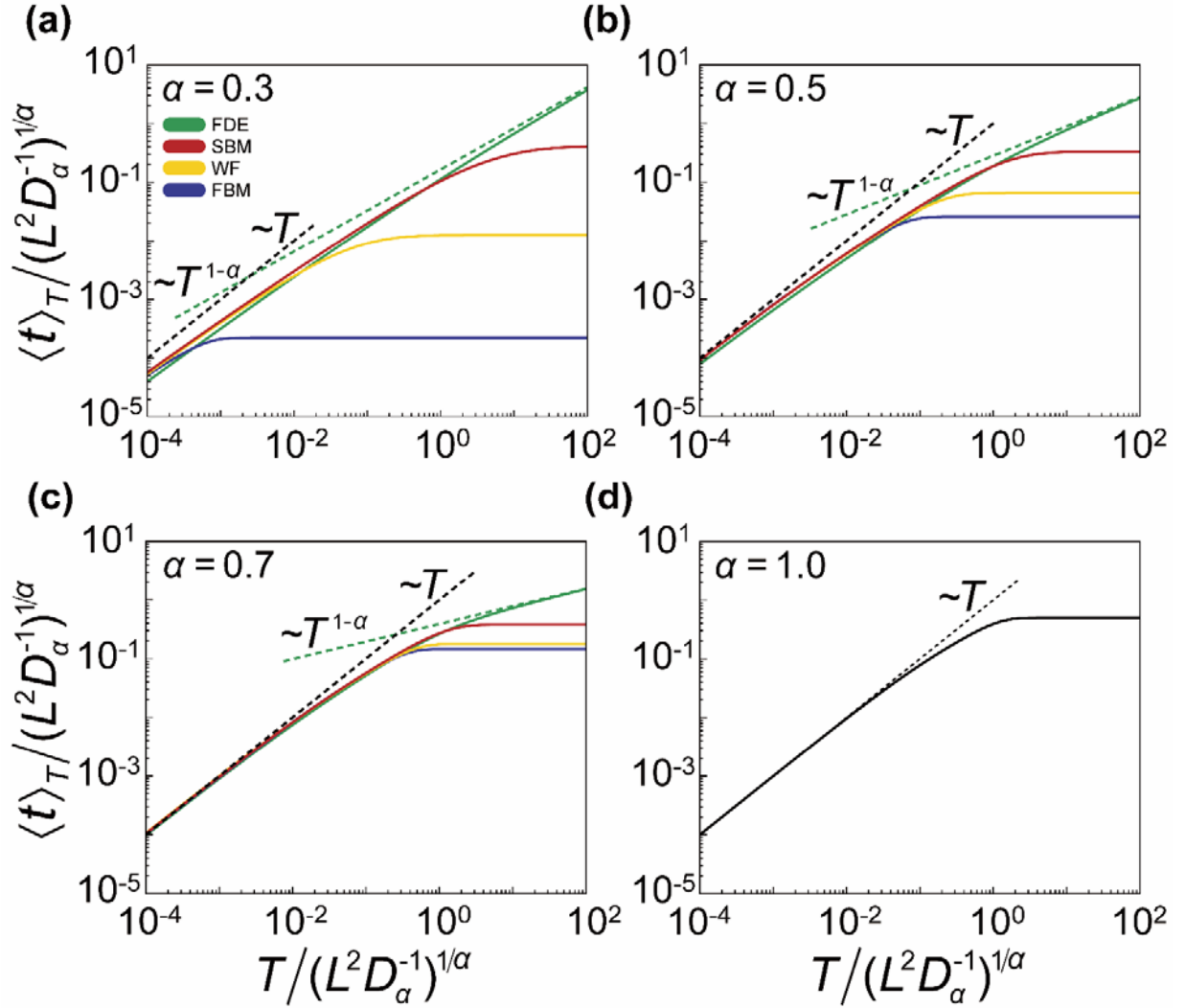

**Figure 1.** The observation time dependence of the $T$-dependent MFPT at various values of $\alpha$. The units of time and length are respectively given by $(L^2 D_\alpha^{-1})^{1/\alpha}$ and $L$. The green, red, yellow, and blue solid lines represent the results for the FDE model, the scaled Brownian motion (SBM), the Wilemski-Fixman (WF) approximation, and the fractional Brownian motion (FBM), respectively. The green dashed line represents the large-$T$ asymptote for the FDE model, which is given by the first term on the right-hand side of equation (14). For the other subdiffusion models except the FDE model, the $T$-dependent MFPTs approach their own MFPTs in the large-$T$ limit. In the small-

$T$ limit, the $T$-dependent MFPTs for all the models approach $T$, which is represented by the black dashed line. When $\alpha = 1$, all the results collapses into the simple diffusion result, which is represented by the black line in (d).

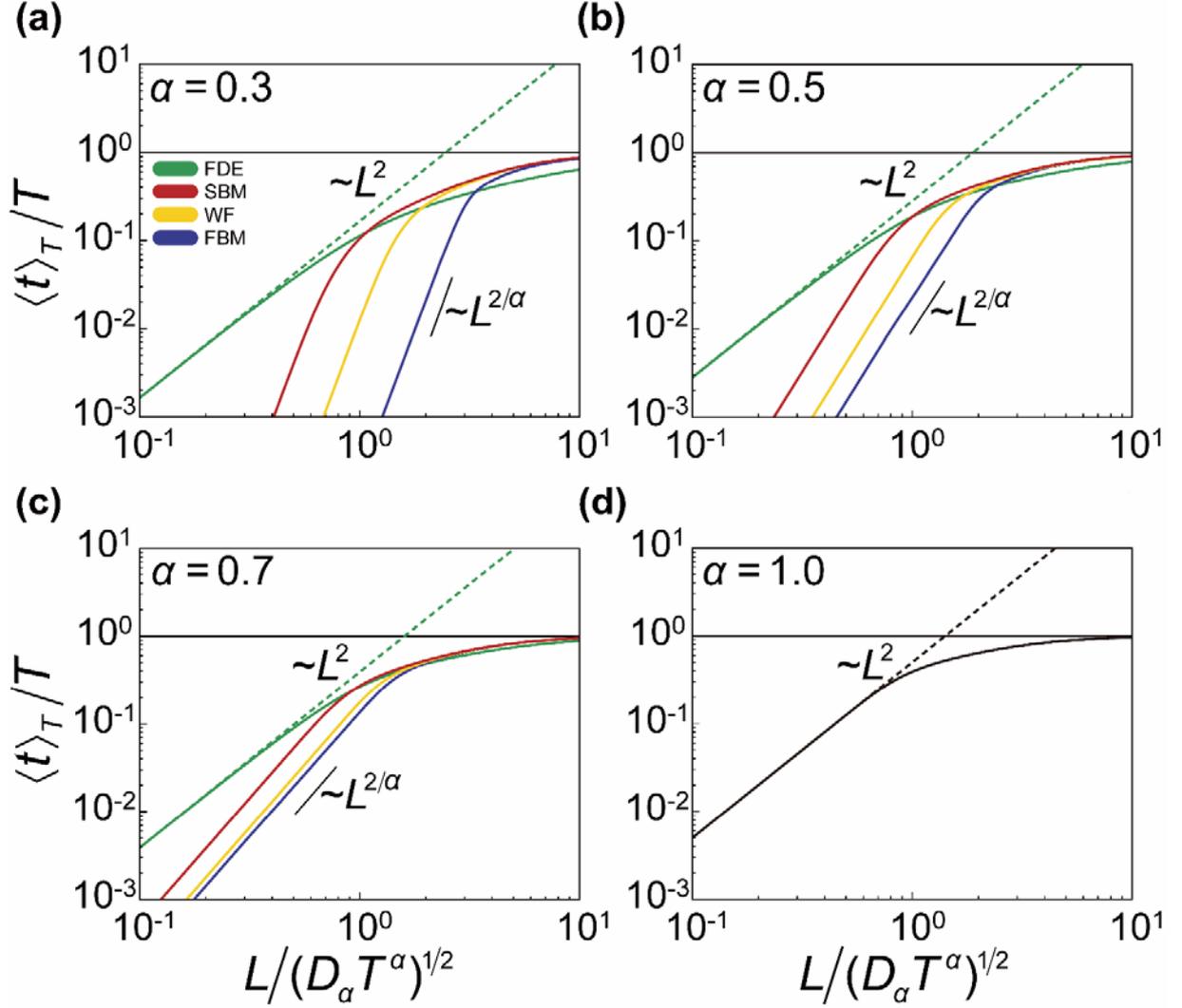

**Figure 2.** The length scale dependence of the $T$-dependent MFPT at various values of $\alpha$. The units of time and length are respectively given by $T$ and $(D_\alpha T^\alpha)^{1/2}$. The green, red, yellow, and blue solid lines represent the results for the FDE model, the scaled Brownian motion (SBM), the Wilemski-Fixman (WF) approximation, and fractional Brownian motion (FBM), respectively. The green dashed line represents the small-$L$ asymptote for the FDE model, which is given by the first term on the right-hand side of equation (14). For the other subdiffusion models except the FDE

model, the T-dependent MFPTs behaves as $\langle t \rangle_T \sim L^{2/\alpha}$ at small *L*. In the large-*L* limit, the *T*-dependent MFPTs for all the models approach *T*. When $\alpha = 1$, all the results collapses into the simple diffusion result, which is represented by the black line in (d).

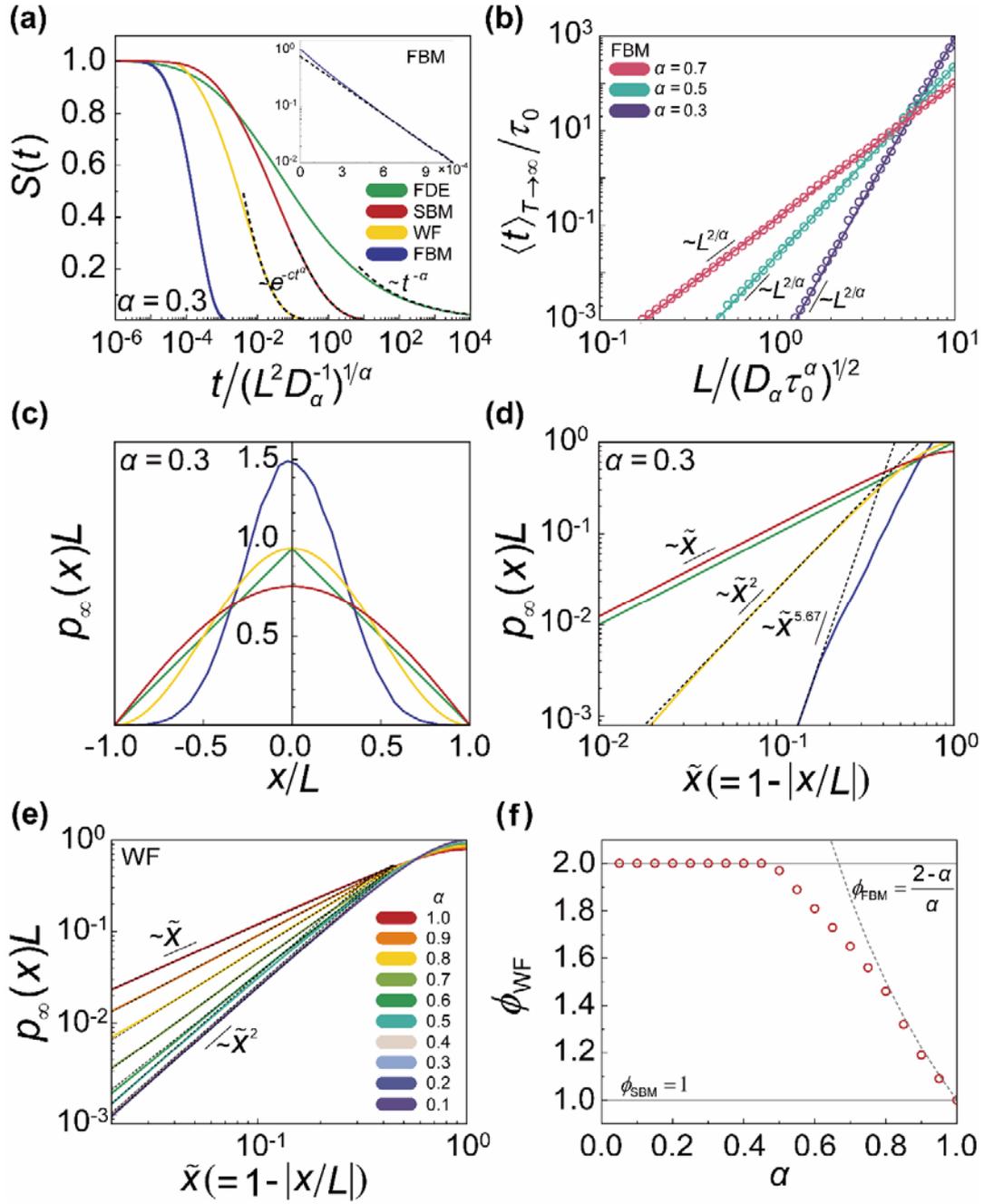

**Figure 3.** Survival probability and normalized displacement distribution. (a) Survival probabilities for four kinds of subdiffusion models at $\alpha = 0.3$. The dashed lines are given as an eye guide to show that the long-time behavior of survival probability follows a stretched exponential decay for the SBM and the WF approximation, and a power-law decay for the FDE model. (inset) Survival probability for the FBM decays exponentially at long times. (b) Length scale dependence of the MFPT for the FBM at various values of $\alpha$. For the FBM, the MFPT scales as $\langle t \rangle_{T \to \infty} \sim L^{2/\alpha}$. $\tau_0$ is an arbitrary time scale chosen as the unit of time. (c) Normalized displacement distributions, $p_\infty(x) [\equiv \lim_{t \to \infty} p(x,t)/S(t)]$, in the long-time limit at $\alpha = 0.3$. (d) Near boundary behavior of $p_\infty(x)$ at $\alpha = 0.3$. $p_\infty(x)$ behaves as $p_\infty(x) \sim \tilde{x}^\phi$ near absorbing boundaries, where $\tilde{x}$ denotes the normalized distance between $x$ and the absorbing boundary, i.e., $\tilde{x} = 1 - |x/L|$. The value of $\phi$ is dependent on the subdiffusion model. (e) Near boundary behavior of $p_\infty(x)$ for the WF approximation at various values of $\alpha$. (f) The $\alpha$-dependence of $\phi$ for the WF approximation. The dotted line represents the $\alpha$-dependence of $\phi$ for the FBM, which is given by $\phi_{FBM} = (2-\alpha)/\alpha$. For the SBM, $\phi$ is just unity, independent of $\alpha$.